\documentstyle[12pt,palatino,epsfig]{article}

\catcode`@=11
\def\citer{\@ifnextchar [{\@tempswatrue\@citexr}{\@tempswafalse\@citexr[]}}
 
\def\@citexr[#1]#2{\if@filesw\immediate\write\@auxout{\string\citation{#2}}\fi
  \def\@citea{}\@cite{\@for\@citeb:=#2\do
    {\@citea\def\@citea{--\penalty\@m}\@ifundefined
       {b@\@citeb}{{\bf ?}\@warning
       {Citation `\@citeb' on page \thepage \space undefined}}%
\hbox{\csname b@\@citeb\endcsname}}}{#1}}
\catcode`@=12

\def\refeq#1{\mbox{eq.~(\ref{#1})}}

\def\reffi#1{\mbox{Fig.~\ref{#1}}}

\def\citere#1{\mbox{Ref.~\cite{#1}}}

\newcommand{\mste}{m_{\tilde{t}_1}}
\newcommand{\mstz}{m_{\tilde{t}_2}}

\newcommand{\MstL}{M_{\tilde{t}_L}}
\newcommand{\MstR}{M_{\tilde{t}_R}}

\newcommand{\Mtlr}{M_{t}^{LR}}

\newcommand{\msq}{m_{\tilde{q}}}

\newcommand{\Pe}{\phi_1}
\newcommand{\Pz}{\phi_2}
\newcommand{\PePz}{\phi_1\phi_2}
\newcommand{\mpe}{m_{\Pe}}
\newcommand{\mpz}{m_{\Pz}}
\newcommand{\mpez}{m_{\PePz}}

\newcommand{\oaas}{{\cal O}(\alpha\alpha_s)}
\newcommand{\cp}{{\cal CP}}

\newcommand{\twol}{two-loop}
\newcommand{\onel}{one-loop}

\newcommand{\MA}{M_A}
\newcommand{\mh}{m_h}

\newcommand{\mt}{m_{t}}

\newcommand{\mgl}{m_{\tilde{g}}}

\newcommand{\Stop}{\tilde{t}}
\newcommand{\StopL}{\tilde{t}_L}
\newcommand{\StopR}{\tilde{t}_R}
\newcommand{\Stope}{\tilde{t}_1}
\newcommand{\Stopz}{\tilde{t}_2}

\newcommand{\tst}{\theta_{\tilde{t}}}

\newcommand{\tsf}{\theta\kern-.20em_{\tilde{f}}}
\newcommand{\tsfp}{\theta\kern-.20em_{\tilde{f}\prime}}
\newcommand{\tsq}{\theta\kern-.15em_{\tilde{q}}}

\newcommand{\VL}{\left( \begin{array}{c}}
\newcommand{\VR}{\end{array} \right)}
\newcommand{\ML}{\left( \begin{array}{cc}}
\newcommand{\MLd}{\left( \begin{array}{ccc}}
\newcommand{\MLv}{\left( \begin{array}{cccc}}
\newcommand{\MR}{\end{array} \right)}

\newcommand{\Tb}{\tan \beta\hspace{1mm}}

\newcommand{\CTb}{\cot \beta\hspace{1mm}}

\newcommand{\tev}{\,\, {\mathrm TeV}}
\newcommand{\gev}{\,\, {\mathrm GeV}}

\newcommand{\BC}{\begin{center}}
\newcommand{\EC}{\end{center}}
\newcommand{\BE}{\begin{equation}}
\newcommand{\EE}{\end{equation}}
\newcommand{\BEA}{\begin{eqnarray}}
\newcommand{\BEAnn}{\begin{eqnarray*}}
\newcommand{\EEA}{\end{eqnarray}}
\newcommand{\EEAnn}{\end{eqnarray*}}

\newcommand{\id}{{\rm 1\kern-.12em
\rule{0.3pt}{1.5ex}\raisebox{0.0ex}{\rule{0.1em}{0.3pt}}}}

\def\hSi{\hat{\Sigma}}

\begin{document}                                                              
\null
\hfill KA-TP-06-1998\\
\null
\hfill hep-ph/9806250\\
\vskip .8cm
%\vspace*{0.8cm}

%
\begin{center}
{\bf
TWO-LOOP RESULTS FOR THE MASSES OF THE\\ 
NEUTRAL CP-EVEN HIGGS BOSONS IN THE MSSM%
\footnote{Talk given by G.~Weiglein at the XXXIIIrd Rencontres de
Moriond, Electroweak Interactions and Unified Theories, Les Arcs,
March 14--21 1997, to appear in the proceedings.}
}\\
\vspace*{0.7cm}

{S.~Heinemeyer, W.~Hollik and G.~Weiglein \\}
{\sl Institut f\"ur Theoretische Physik,
     Universit\"at Karlsruhe, \\
     D--76128 Karlsruhe, Germany \\}
\end{center}

\vspace{.7cm}

\begin{abstract}
\noindent
Diagrammatic \twol\ results are presented for the leading QCD corrections 
to the masses of the neutral $\cp$-even Higgs bosons in the Minimal
Supersymmetric Standard Model (MSSM). The results are valid for
arbitrary values of the parameters of the Higgs and scalar top sector
of the MSSM.
%and give rise to a considerable reduction of the mass of
%the lightest Higgs boson compared to its one-loop value.  
Their impact on a precise prediction for the mass of the lightest Higgs
boson is briefly discussed.
\end{abstract}

%\newpage

%%%%%%%%%%%%%%%%%%%%%%%%%%%%%%%%%%%%%%%%%%%%%%%%%%%%%%%%%%%%%%
%%%%%%%%%%%%%%%%%%%%%%%%%%%%%%%%%%%%%%%%%%%%%%%%%%%%%%%%%%%%%%

%The Minimal Supersymmetric %extension of the Standard Model (MSSM)
The MSSM
offers a theoretically very appealing extension of the Standard 
Model (SM) of the electroweak and strong interactions. While for 
the predicted superpartners of the SM particles present-day
accelerators can cover only a very limited part of the MSSM parameter
space, the Higgs sector of the model provides the opportunity for a
stringent direct test in requiring the existence of a light neutral
Higgs boson. At tree level its mass, $\mh$, is predicted to be lower
than the one of the Z~boson. However, the \onel\ corrections are
known to be huge~\cite{mhiggs1l,mhoneloop} and shift the upper bound to about 
$150 \gev$. Beyond \onel\ order renormalization group methods have been
applied in order to include higher-order leading and 
next-to-leading logarithmic
contributions~\cite{mhiggsRG1,mhiggsRG1a,mhiggsRG2}. A diagrammatic
calculation is available for the dominant \twol\
contributions in the limiting case of vanishing $\Stop$-mixing and
infinitely large $\MA$ and $\Tb$~\cite{hoanghempfling}. 
These results indicate that the
\twol\ corrections considerably reduce the predicted value of $\mh$. 
A precise prediction for $\mh$ in terms of the relevant SUSY parameters
is crucial in order to determine the discovery and exclusion potential 
of LEP2 and the upgraded Tevatron and also for physics at the LHC,
where eventually a high-precision measurement of $\mh$ might be
possible. 

Diagrammatic \twol\ results 
%for general parameters of the MSSM Higgs sector and taking 
which take into account virtual particle effects without
restrictions of their masses and mixing are therefore very desirable.
We have performed a Feynman diagrammatic calculation of the leading
\twol\ QCD corrections to the masses of the neutral $\cp$-even Higgs
bosons in the MSSM~\cite{mhiggslett}. The calculation has been
performed in the on-shell scheme. The results are valid for arbitrary
values of the parameters of the Higgs and scalar top sector of the
MSSM.

%%%%%%%%%%%%%%%%%%%%%%%%%%%%%%%%%%%%%%%%%%%%%%%%%%%%%%%%%%%%%%
%%%%%%%%%%%%%%%%%%%%%%%%%%%%%%%%%%%%%%%%%%%%%%%%%%%%%%%%%%%%%%

The Higgs sector of the MSSM contains two Higgs doublets,
\BEA
H_1 = \VL H_1^1 \\ H_1^2 \VR = \VL v_1 + (\phi_1^{0} + i\chi_1^{0})
                                 /\sqrt2 \\ \phi_1^- \VR , \quad
H_2 = \VL H_2^1 \\ H_2^2 \VR =  \VL \phi_2^+ \\ v_2 + (\phi_2^0 
                                     + i\chi_2^0)/\sqrt2 \VR,
\label{eq:hidoubl}
\EEA
and can be described using the two input parameters 
$\Tb = v_2/v_1$ and $\MA^2 = -m_{12}^2(\Tb+\CTb)$,
where $\MA$ is the mass of the $\cp$-odd A~boson.

%At tree level the mass matrix of the neutral $\cp$-even Higgs bosons
%is given in the $\phi_1-\phi_2$ basis 
%in terms of $\MZ$ and $\MA$ through
%\BEA
%M_{\rm Higgs}^{2, {\rm tree}} = \ML \mpe^2 & \mpez^2 \\ 
%                           \mpez^2 & \mpz^2 \MR =
%\ML \MA^2 \SQb + \MZ^2 \CQb & -(\MA^2 + \MZ^2) \Sb \Cb \\
%    -(\MA^2 + \MZ^2) \Sb \Cb & \MA^2 \CQb + \MZ^2 \SQb \MR.
%\EEA

The tree-level mass predictions are affected by large corrections at one-loop
order through terms proportional to $G_F \mt^4 \ln
(\mste \mstz/\mt^2)$~\cite{mhiggs1l}. These 
dominant one-loop contributions can be obtained by evaluating the
contribution of the $t-\Stop$-sector to the $\phi_{1,2}$ self-energies at zero
external momentum from the Yukawa part of the theory 
(neglecting the gauge couplings). Accordingly, the one-loop corrected 
Higgs masses are derived by diagonalizing the mass matrix (given 
in the $\phi_1-\phi_2$ basis; $\mpe^2, \mpez^2, \mpz^2$ are the
tree-level entries)
\BE
M^2_{\rm Higgs}%^{\rm 1-loop}
%= \ML \mpe^2 - \hSi_{\Pe}(0) & \mpez^2 - \hSi_{\PePz}(0) \\
%     \mpez^2 - \hSi_{\PePz}(0) & \mpz^2 - \hSi_{\Pz}(0) \MR,
= \VL \mpe^2 - \hSi_{\Pe}(0)\;\;\;\;\;\; \mpez^2 - \hSi_{\PePz}(0) \\
     \mpez^2 - \hSi_{\PePz}(0)\;\;\;\;\;\; \mpz^2 - \hSi_{\Pz}(0) \VR .
\label{higgsmassmatrixnondiag}
\EE
Here the $\hSi$ denote the Yukawa contributions of the $t-\Stop$-sector
to the renormalized \onel\ $\phi_{1,2}$ self-energies. 
By comparison with the full \onel\ result~\cite{mhoneloop} it has been
shown that these contributions indeed contain the bulk of the \onel\
corrections. 
They typically approximate the full \onel\ result up to about 5~GeV.
%Their deviation from the full \onel\ result is of the order of 5~GeV.
%For small $\Stop$-mixing and $\Tb > 2$ they approximate
%the full \onel\ result up to 2--3~GeV, while for larger
%$\Stop$-mixing agreement within up to about $10 \gev$ is found.
%It has been shown
%that these contributions approximate the full \onel\ result for small
%$\Stop$-mixing and $\Tb > 2$ up to $2-3
%\gev$~\cite{mhiggs1lstop2}, although a deviation of up to $10 \gev$ is
%possible for larger $\Stop$-mixing.

In order to derive the leading two-loop contributions to the masses of
the neutral $\cp$-even Higgs bosons we have evaluated the QCD
corrections to \refeq{higgsmassmatrixnondiag}, which because of the
large value of the strong coupling constant are expected to be the most
sizable ones (see also \citere{hoanghempfling}). This requires the
evaluation of the renormalized $\phi_{1,2}$ self-energies at the
\twol\ level. 
The renormalization has been performed in the on-shell scheme.
The counterterms in the Higgs sector are derived from the
Higgs potential. The renormalization conditions for
the tadpole counterterms have
been chosen in such a way that they cancel the tadpole contributions in
one- and \twol\ order. 
The renormalization in the $t-\Stop$-sector has
been performed in the same way as in \citere{drhosuqcd}.
For the present calculation the \onel\ counterterms $\delta m_t$,
$\delta m_{\Stope}$, $\delta m_{\Stopz}$ for the top-quark
and scalar top-quark masses and $\delta \tst$ for the mixing angle
contribute, which enter via the subloop renormalization. The appearance
of the %scalar top 
$\Stop$ mixing angle $\tst$ reflects the fact that the current
eigenstates, $\StopL$ and $\StopR$, mix to give the mass eigenstates
$\Stope$ and $\Stopz$. 
Since the non-diagonal entry in the scalar
quark mass matrix is proportional to the
quark mass the mixing is particularly important in the case of the third
generation scalar quarks.
%
%The renormalized self-energies have the following structure:
%\BE
%\label{P1serenb}
%\hSi_s(0) = \Sie_s(0) + \Siz_s(0)
%- \de V_s^{(1)} - \de V_s^{(2)},
%\EE
%where $s = \phi_1, \phi_2, \phi_1 \phi_2$. $\Sie_s$ and $\Siz_s$ denote
%the unrenormalized self-energies at the one- and \twol\ level, and
%$\de V^{(1)}_s$ and $\de V^{(2)}_s$ are the one- and \twol\ counterterms 
%derived from the Higgs potential~\cite{mhiggslett}. 
%The counterterms read:
%\BEA
%\label{P1potctMW0Yuk1l}
%\de V_{\Pe}^{(i)} &=& 
%             + \de\MA^{2(i)} \SQb
%             - \de t_1^{(i)} \frac{e\, \Cb}{2 \MW \sw} (1 + \SQb) 
%           + \de t_2^{(i)} \frac{e}{2 \MW \sw} \CQb \Sb, \\
%\label{P2potctMW0Yuk1l}
%\de V_{\Pz}^{(i)} &=& 
%             + \de\MA^{2(i)} \CQb
%             - \de t_2^{(i)} \frac{e\, \Sb}{2 \MW \sw} (1 + \CQb) 
%           + \de t_1^{(i)} \frac{e}{2 \MW \sw} \SQb \Cb, \\
%\label{P1P2potctMW0Yuk1l}
%\de V_{\PePz}^{(i)} &=& 
%               - \de\MA^{2(i)} \Sb\Cb
%               - \de t_1^{(i)} \frac{e}{2 \MW \sw} \SDb 
%             - \de t_2^{(i)} \frac{e}{2 \MW \sw} \CDb, 
%\EEA
%with $\de t_a^{(i)} = - T_a^{(i)}$, where $T_a^{(i)}$ denotes the tadpole
%contribution, $\de t_a^{(i)}$ is the corresponding counterterm, 
%and $\de\MA^{2(i)} = \Si_A^{(i)}(0)$ ($i = 1,2$).
%
In deriving our results we have made strong use of computer-algebra
tools~\cite{compalg}. 
%Our results have been derived by the wide usage of computer-algebra
%programs. 
%The package \fa~\cite{feynarts} (in which the relevant part
%of the MSSM has been implemented) has been applied to generate the
%Feynman amplitudes and the counterterm contributions. For evaluating the
%amplitudes the package \tc~\cite{twocalc} has been used.
The calculations have been performed using Dimensional Reduction~\cite{dred},
which preserves the 
%(DRED)~\cite{dred}, which is necessary in order to preserve the
relevant SUSY relations. 
%Application of Dimensional Regularization
%(DREG)~\cite{dreg}, on the other hand, does not lead to a finite result.
%The same observation has also been made in~\citere{hoanghempfling}.
As results we have obtained analytical expressions for the \twol\
%Our results for the \twol\ $\phi_{1,2}$ self-energies are given 
$\phi_{1,2}$ self-energies in terms of the
SUSY parameters $\Tb$, $\MA$, $\mu$, $\mste$, $\mstz$, $\tst$, and $\mgl$. 
%In the general case the results
%are by far too lengthy to be given here
%explicitly. In the special case of vanishing mixing in the 
%$\Stop$-sector, $\mu = 0$, and $\mste = \mstz = \mst$, a relatively 
%compact expression can be derived, which has been given in
%\citere{mhiggslett}.

Inserting the \onel\ and \twol\ $\phi_{1,2}$ self-energies into 
\refeq{higgsmassmatrixnondiag}, the predictions for the masses of the
neutral $\cp$-even Higgs bosons are derived by diagonalizing the 
\twol\ mass matrix.
For the numerical evaluation we have chosen two values for $\Tb$ which
are favored by SUSY-GUT scenarios~\cite{su5so10}: $\Tb = 1.6$ for
the $SU(5)$ scenario and $\Tb = 40$ for the $SO(10)$ scenario. 
%Other parameters are $\MZ = 91.187 \gev, \MW = 80.375 \gev, 
%G_F = 1.16639 \, 10^{-5} \gev^{-2}, \als = 0.1095$, and $\mt = 175 \gev$. 
%For the figures below we have furthermore
%chosen $\mu = - 200 \gev, \MA = 500 \gev$, and $\mgl = 500 \gev$ as typical
%values. 
The scalar top masses and the mixing angle are derived from the
parameters $M_{{\tilde t}_L}$, $M_{{\tilde t}_R}$ and $\Mtlr$ of the 
$\Stop$ mass matrix, where $\Mtlr = A_t - \mu \cot\beta$ 
(our conventions are the same as in \citere{drhosuqcd}). 
In the figures below we have chosen \mbox{$\msq \equiv \MstL = \MstR$}.

\begin{figure}[htb]
\vspace{-.5cm}
\begin{center}
\mbox{
\psfig{figure=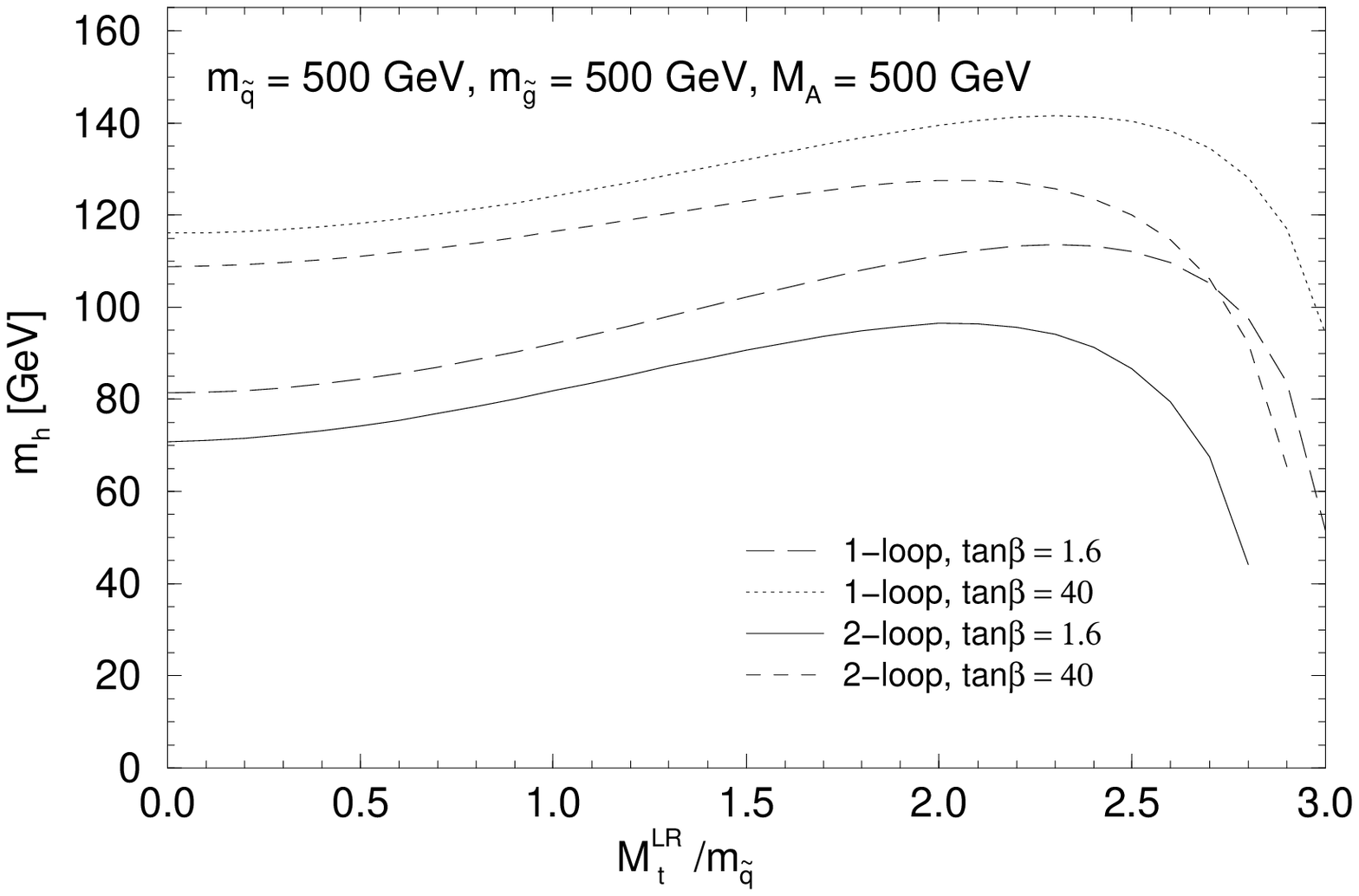,width=7cm,height=6.5cm,
                      bbllx=150pt,bblly=130pt,bburx=450pt,bbury=420pt}}
\end{center}
\caption[]{One- and \twol\ results for $\mh$ as a function of
$\Mtlr/\msq$ for two values of $\Tb$.}
\label{fig:plot1}
\end{figure}

The plot in Fig.~\ref{fig:plot1} shows $\mh$
as a function of $\Mtlr/\msq$, where $\msq$
%, the common scalar breaking parameter in the $\Stop$ mass matrix, 
is fixed to $500 \gev$. A minimum is reached for  
$\Mtlr = 0 \gev$ which we refer to as `no mixing'. A maximum in the
\twol\ result for $\mh$ is reached for about $\Mtlr/\msq \approx 2$ 
in the $\Tb = 1.6$ scenario as well as in the $\Tb = 40$ scenario. 
This case we refer to as `maximal mixing'. Note that the maximum is
shifted compared to its \onel\ value of about $\Mtlr/\msq \approx 2.4$.

%%%%%%%%%%%%%%%%%%%%%%%%%%%%%%%%%%%%%%%%%%%%%%%%%%%%%%%%%%%%%%
%%%%%%%%%%%%%%%%%%%%%%%%%%%%%%%%%%%%%%%%%%%%%%%%%%%%%%%%%%%%%%

\begin{figure}[htb]
\begin{center}
\hspace{1em}
\mbox{
\psfig{figure=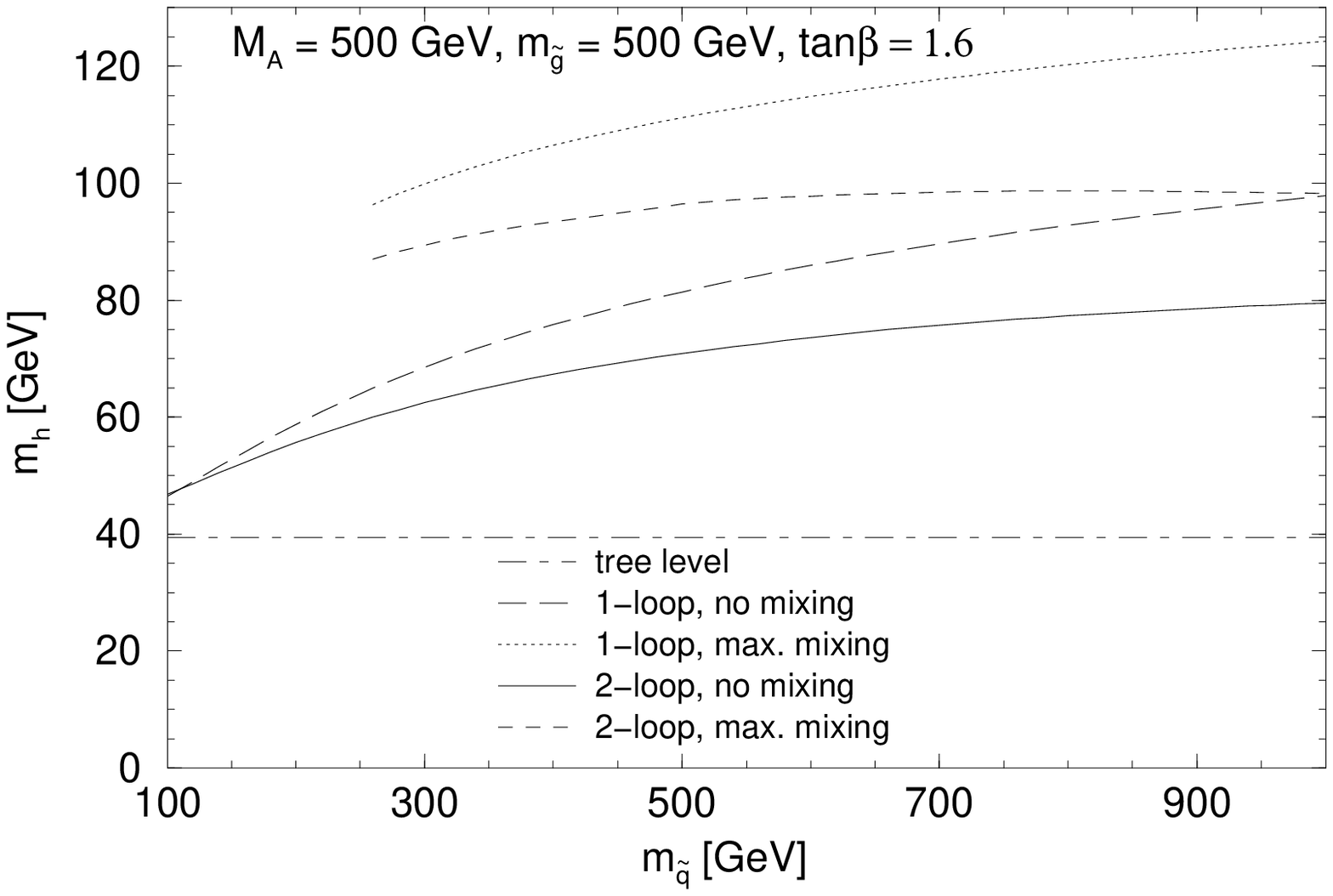,width=5.3cm,height=8cm,
                      bbllx=150pt,bblly=100pt,bburx=450pt,bbury=420pt}}
\hspace{7.5em}
\mbox{
\psfig{figure=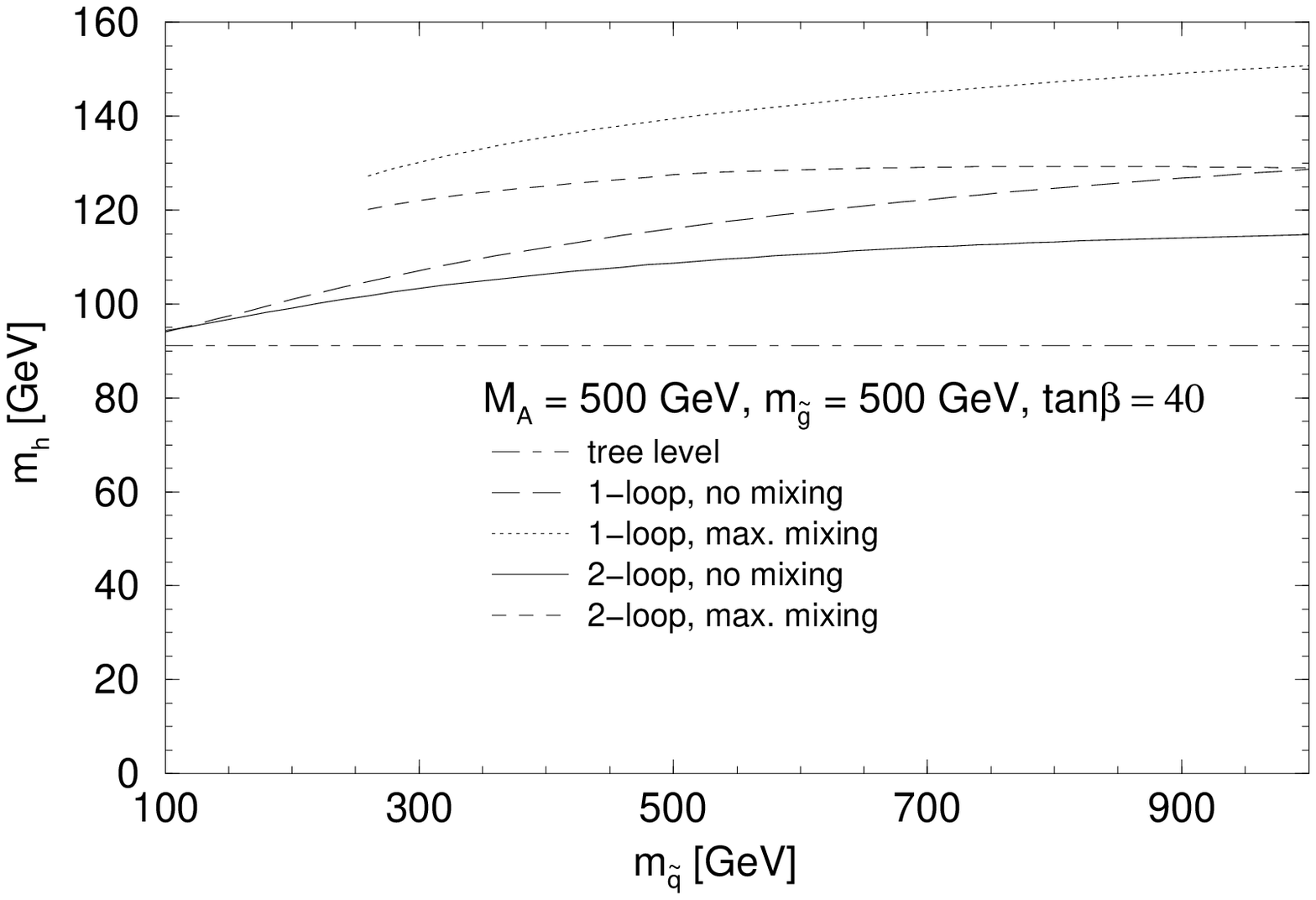,width=5.3cm,height=8cm,
                      bbllx=150pt,bblly=100pt,bburx=450pt,bbury=420pt}}
\end{center}
\caption[]{%$\mh$ in the $SU(5)$ scenario, 
The mass of the lightest Higgs boson for the two scenarios with
$\Tb = 1.6$ and $\Tb = 40$. 
The tree-, the one- and the \twol\ results for $\mh$ are shown 
as a function of $\msq$ for the no-mixing and the maximal-mixing case.}
\label{fig:plot2}
\end{figure}

In Fig.~\ref{fig:plot2} the two scenarios with $\Tb = 1.6$ and
$\Tb = 40$ are analyzed.
The tree-level, the \onel\ and the \twol\ results for $\mh$
are shown as a function of $\msq$ for no mixing and maximal mixing
(the curves for the maximal-mixing case start at higher values of 
$\msq$ than those for the no-mixing case, since below these values of
$\msq$ the resulting $\Stop$-masses are unphysical or experimentally
excluded). The plots show that the \onel\ on-shell result for $\mh$ is
in general considerably reduced by the inclusion of the \twol\
corrections. For the low-$\Tb$ scenario the difference between the
\onel\ and \twol\ result amounts to up to about $18 \gev$ for $\msq = 1
\tev$ in the no-mixing case, and up to about $25 \gev$ for $\msq = 1
\tev$ in the maximal-mixing case. For the high-$\Tb$ scenario the
reduction of the \onel\ result is slightly smaller than for $\Tb = 1.6$.
The variation of our results with $\mgl$ is of the order of few GeV.

%%%%%%%%%%%%%%%%%%%%%%%%%%%%%%%%%%%%%%%%%%%%%%%%%%%%%%%%%%%%%%
%%%%%%%%%%%%%%%%%%%%%%%%%%%%%%%%%%%%%%%%%%%%%%%%%%%%%%%%%%%%%%

%We have compared our results with the results obtained in
%\citere{hoanghempfling} in the case of no $\Stop$-mixing and $\MA \to \infty,
%\Tb \to \infty$ and have checked analytically that in the limiting case
%$\mste = \mstz = \mgl \gg \mt$ we recover the corresponding formula
%given in \citere{hoanghempfling}.

Supplementing our results for the leading $\oaas$ corrections with the
leading higher-order Yukawa term of ${\cal O}(\alpha^2 \mt^6)$ given in
\citere{mhiggsRG1a} leads to an increase in the prediction of $\mh$ of
up to about 3~GeV.
A similar shift towards higher values of $\mh$ emerges if at
the two-loop level the running top-quark mass, $\mt = 166.5 \gev$, is
used instead of the pole mass, $\mt = 175 \gev$, thus taking into
account leading higher-order effects beyond the \twol\ level. 
We have compared our results with
the results of a renormalization group improvement of the leading
one-loop contributions given in \citere{mhiggsRG2}. We find good
agreement for the case of no $\Stop$-mixing, while for larger
$\Stop$-mixing sizable deviations exceeding 5~GeV occur. In
particular, the value of $\Mtlr/\msq$ for which $\mh$ becomes maximal
is shifted from $\Mtlr/\msq \approx 2.4$ in the \onel\ case to 
$\Mtlr/\msq \approx 2$ when our
diagrammatic \twol\ results are included (see \reffi{fig:plot1}).
In the results based on renormalization group 
methods~\cite{mhiggsRG1,mhiggsRG2}, on the other hand, the maximal
value of $\mh$ is obtained for $\Mtlr/\msq \approx 2.4$, i.e.\ at the
\onel\ value.

In summary, we have diagrammatically calculated the leading $\oaas$
corrections to the masses of the neutral $\cp$-even Higgs bosons in 
the MSSM. 
We have applied the on-shell scheme and have imposed no restrictions on
the parameters of the Higgs and scalar top sector of the model. 
The \twol\ correction leads to a considerable reduction of the
prediction for the mass of the lightest Higgs boson compared to the 
\onel\ value. The reduction turns out to be particularly important for
low values of $\Tb$. Compared to the results obtained via
renormalization group methods we find sizable deviations for large 
mixing in the $\Stop$-sector.\\[-.7em]
%The leading \twol\ contributions presented here can directly be combined
%with the complete \onel\ results in the on-shell scheme~\cite{mhoneloop}.
%A discussion of the corresponding results will be given in a
%forthcoming paper, where also a more detailed comparison with the
%results based on renormalization group methods will be pursued.\\

G.W.\ thanks the organizers of the ``XXXIIIrd Rencontres de Moriond''
for the pleasant atmosphere enjoyed at Les Arcs.

%%%%%%%%%%%%%%%%%%%%%%%%%%%%%%%%%%%%%%%%%%%%%%%%%%%%%%%%%%%%%%
%%%%%%%%%%%%%%%%%%%%%%%%%%%%%%%%%%%%%%%%%%%%%%%%%%%%%%%%%%%%%%

\end{document}